\documentclass[aps,prl,reprint,superscriptaddress]{revtex4-2}

\usepackage{graphicx}
\usepackage{dcolumn}
\usepackage{bm}
\usepackage{color}
\usepackage{xcolor}
\usepackage{epstopdf}
\usepackage{gensymb}
\usepackage{ulem}
\usepackage{float}
\usepackage{xr}
\usepackage{sidecap}
\usepackage{mathrsfs}
\usepackage{amsmath}
\usepackage{amsthm}
\usepackage{enumerate}
\usepackage{soul}
\usepackage{amssymb}
\usepackage{units}
\usepackage{lipsum}
\usepackage{comment}

\begin{document}
\title{Inerter-based Elastic Metamaterials for Band Gap at Extremely Low Frequency}

\author{Faisal Jamil}
\affiliation{Department of Mechanical Engineering, University of Utah, Salt Lake City, UT, USA}

\author{Fei Chen}
\affiliation{Department of Mechanical Engineering, University of Utah, Salt Lake City, UT, USA}

\author{Bolei Deng}
\affiliation{Computer Science and Artificial Intelligence Laboratory, Department of Electrical Engineering and Computer Science, and Department of Mechanical Engineering, Massachusetts Institute of Technology, Cambridge, MA, USA}

\author{Robert G. Parker}
\affiliation{Department of Mechanical Engineering, University of Utah, Salt Lake City, UT, USA}

\author{Pai Wang}%
\affiliation{Department of Mechanical Engineering, University of Utah, Salt Lake City, UT, USA}
\thanks{pai.wang@utah.edu}

\begin{abstract}
We reveal the unique and fundamental advantage of inerter-based elastic metamaterials by a comparative study among different configurations. When the embedded inerter is connected to the matrix material on both ends, the metamaterial shows definite superiority in forming a band gap in the ultra-low frequency - equivalently the ultra-long wavelength - regime, where the unit cell size can be four or more orders of magnitude smaller than the operating wavelength. In addition, our parametric studies in both one and two dimensions pave the way towards designing next-generation metamaterials for structural vibration mitigation.\\
\end{abstract}
\maketitle

\begin{figure}[b!]
\includegraphics[width=0.5\textwidth]{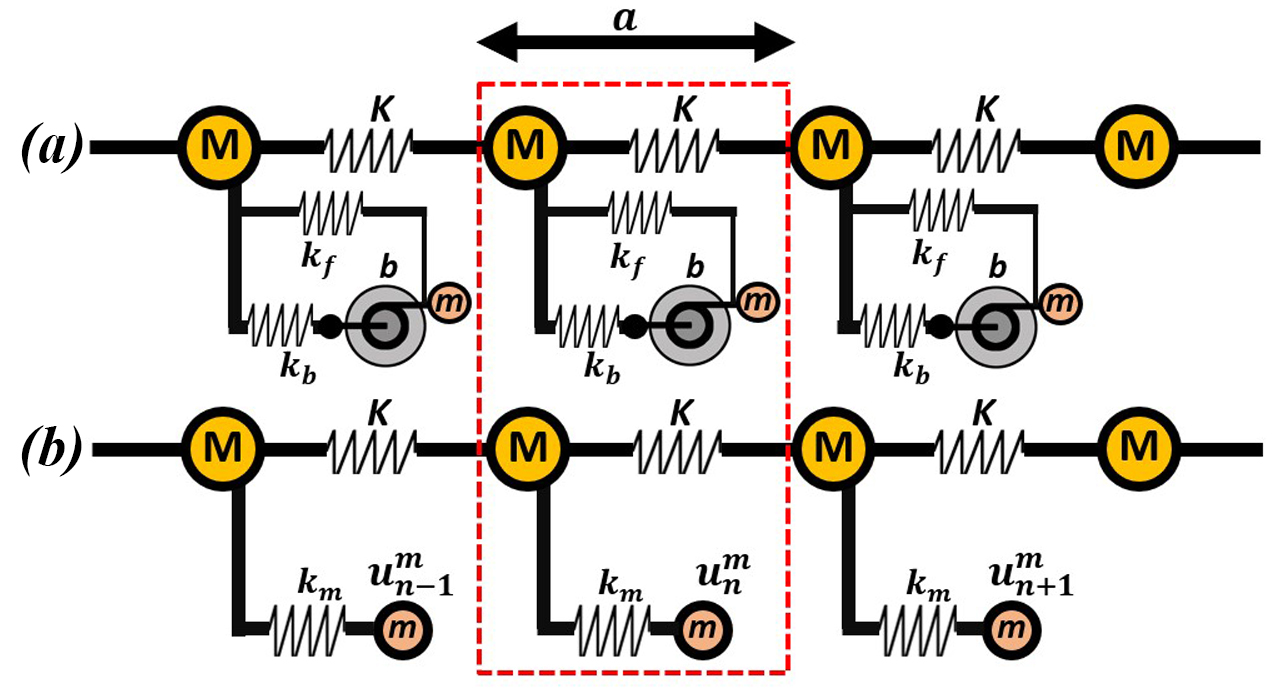}
\caption{\label{fig:para} Discrete models: (a) Metamaterial with inerter-mass-resonators; (b) Metamaterial with mass-resonators;.}
\end{figure}

Mitigation of low-frequency vibrations has long been a major challenge. One promising research direction points to architected materials 
- widely referred to as acoustic or elastic metamaterials~\cite{maldovan2009periodic,hussein2014dynamics,mu2020review,liu2021review,ji2021vibration,Laghfiri_2022}. They can exhibit a phononic band gap, i.e., a range of frequencies in which no vibration can propagate. While many recent studies attempted to demonstrate low-frequency band gaps~\cite{Krodel_2015,Achaoui_2016,Oh_2017,Fang_2017,Krushynska_2018,Muhammad_2019,Chen_2019,D_Alessandro_2019,Sun_2020,Zeng_2020,Ruan_2021,Chen_2021,Lin_2021,Zhang_2021,Huang_2021,Wang_2021,Muhammad_2021,Zhang_2021,Jiang_2021,Zhu_2021,Bae_2022,Li_2022,vo2022reinvestigation,zi2022low,lin2022low,zeng2022seismic,yan2022band}, there is no consensus on which frequency ranges should be called ``low" or ``ultra-low''. The exact meaning of low frequency varies from a fraction of one Hz~\cite{Muhammad_2019,Huang_2021}, to several Hz~\cite{Krodel_2015,Achaoui_2016,Zhang_2021}, and up to many kHz~\cite{Krushynska_2018,Muhammad_2021,vo2022reinvestigation}. The word ``low" is a relative concept that depends on application-specific scenarios.\\
\indent To facilitate a generally meaningful discussion and a fair comparison among different systems and configurations, here we focus on a universal and dimensionless frequency for all vibro-elastic metamaterials: $f = a / \lambda$, where $a$ denotes the size of a metamaterial unit, and $\lambda$ is the operating wavelength. All scattering-based band gaps in phononic crystals~\cite{sigalas1992elastic,kushwaha1993acoustic} are at the order of $f = a/\lambda \sim 1$. In contrast, locally resonant metamaterials embedded with mass-resonators~\cite{liu2000locally,wang2014harnessing} usually exhibit band gaps at a much lower frequency range of $f = a/\lambda \sim 10^{-2}$ to $10^{-3}$.\\
\begin{figure}[b!]
\includegraphics[width=0.5\textwidth]{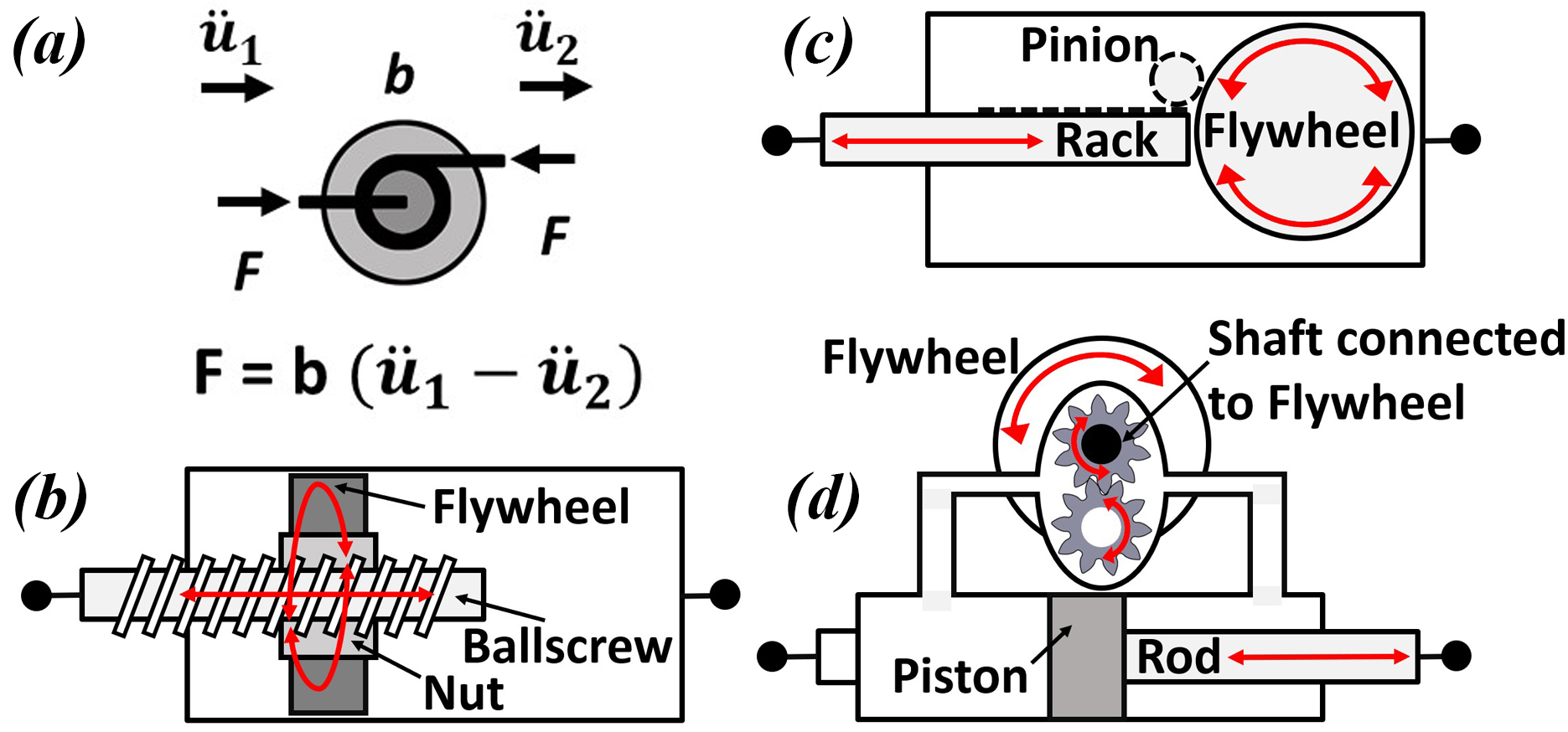}
\caption{\label{fig:inerters} Conceptual schematics of inerters: (a) Abstract symbol and force response. (b) Ball-screw inerter. (c) Rack-and-pinion inerter. (d) Hydraulic inerter.}
\end{figure}
\indent In this Letter, we demonstrate the unique capability of inerter-based metamaterials in forming band gaps at the ultra-low dimensionless frequencies, where $f = a/\lambda \sim 10^{-4}$. The key component is the inerter, a two terminal mechanical device offering a \textit{frequency-independent} inertia much larger than its own physical mass~\cite{chen2009missing,SI}. As illustrated in Fig.\,\ref{fig:inerters}, this is possible because the inerter couples linear relative motions between its two ends to the rotation of a flywheel. The flywheel moment of inertia can be amplified to produce a large inertial effect. The use of rotational motion also makes it possible for the device to be compact. Like springs and dampers, the inerter is a passive device without the need of any active control. As shown in Fig.\,\ref{fig:inerters}(a), the inerter's behavior is characterized by the response force, $F = b (\ddot{u}_1 - \ddot{u}_2)$, where $\ddot{u}_1$ and $\ddot{u}_2$ are the accelerations at the two terminals. The constant $b$ is called the inertance, which has the same unit as mass. The performance attributes of inerters have been experimentally verified with ball-screw designs~\cite{papageorgiou2005laboratory, chen2009missing, papageorgiou2009experimental, shen2016improved, sun2016performance, yuehao2021modeling, yuehao2022study}, rack-pinion designs~\cite{papageorgiou2005laboratory, chen2009missing, papageorgiou2009experimental, wang2014design}, and hydraulic designs~\cite{wang2011designing, nakaminami2017dynamic}. Their basic structures are shown in Figs.\,\ref{fig:inerters}(b), \ref{fig:inerters}(c) and \ref{fig:inerters}(d), respectively. In particular, the hydraulic inerter design benefits from 
the small physical mass of nearly incompressible fluid that fills the inerter, so that it can produce an inertance that is $1.5\times10^6$ times larger than its own the physical mass \cite{nakaminami2017dynamic}.\\
\begin{figure}[h]
\includegraphics[width=0.5\textwidth]{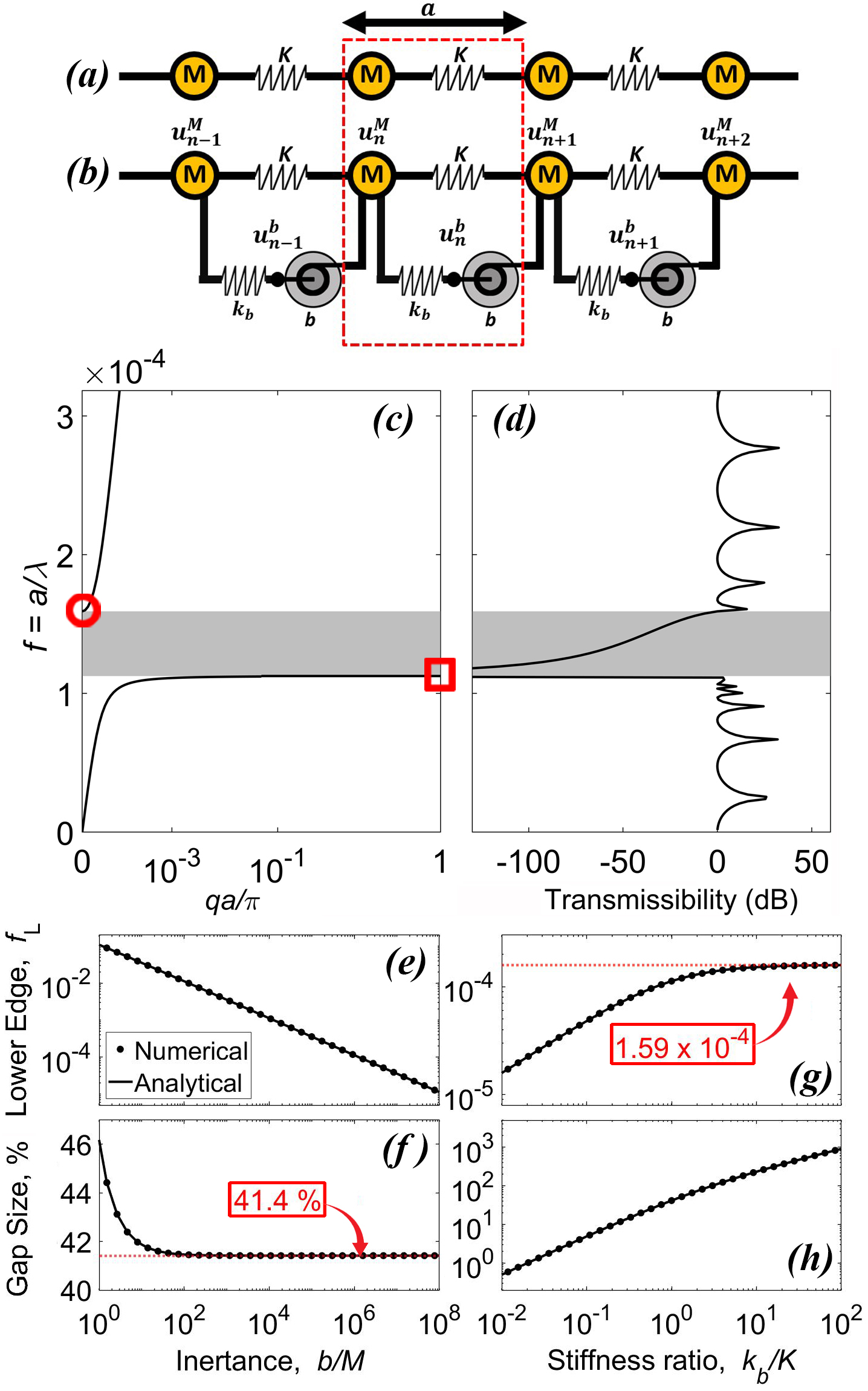}
\caption{\label{fig:meta} Discrete models and band gaps: (a) Abstraction of any matrix material as a spring-mass chain; (b) Inerter-based metamaterial; (c) Dispersion curves and band gap (shaded region) of inerter-based metamaterial shown in (b) with $b/m_b = b/M = 10^6$ and $k_b / K = 1$; and (d) Finite-chain response of inerter-based metamaterial with 1000 unit cells. The transmissibility is calculated as the ratio between output and input amplitudes. Parametric studies on the band gap lower edge frequency and relative size: (e) and (f) show the effects of change in the inertance ratio, $\mu_b=b/M$, with fixed $\kappa_b=k_b/K=1$. (g) and (h) show the effects of change in the stiffness ratio, $\kappa_b=k_b/K$, with fixed $\mu_b =b/M = 10^{6}$.
}\end{figure}
\indent While there have been several pioneering attempts to incorporate inerters into metamaterial designs~\cite{kulkarni2016longitudinal,al2018dispersion,fang2018band,madhamshetty2019extraordinary,sun2019bandgap,cajic2021tuning,liu2022enhanced}. A critical study to identify and overcome the fundamental hurdles is still missing. With theoretical and numerical analyses of different designs, we investigate the basic challenges and offer a road map to realize vibration-band-gap metamaterials with unit cells in the ultra-deep sub-wavelength scale of $a/\lambda \sim 10^{-4}$.\\
\indent To start, we model any matrix material or base structure of metamaterials as a spring-mass chain with stiffness $K$ and point mass $M$, as shown in Fig.\,\ref{fig:meta}(a). This model is valid since we aim at the long-wavelength limit of $\lambda \gg a$, where the discrete nature of the main chain has negligible impact on the results. With this setup, we can normalize all metamaterial dispersion relations according to the main-chain wave speed in the long-wavelength limit, $c = a \sqrt{K/M}$, so that all band gap frequencies are non-dimensionalized~\cite{maldovan2009periodic, SI} as $f = \omega a / (2 \pi c) = a / \lambda$, where $\omega$ is the dimensional angular frequency in metamaterial dispersion relations.\\
\indent We analyze three types of metamaterial designs with: embedded inerters~\cite{kulkarni2016longitudinal, al2018dispersion, sun2019bandgap, cajic2021tuning}, inerter-mass-resonators~\cite{kulkarni2016longitudinal, al2018dispersion, madhamshetty2019extraordinary, sun2019bandgap, fang2018band, liu2022enhanced, wang2022frequency}, and traditional mass-resonators~\cite{liu2000locally,wang2014harnessing}, as shown in Figs.\,\ref{fig:meta}(b), \ref{fig:para}(a) and \ref{fig:para}(b), respectively.
Applying the Bloch theorem~\cite{hussein2014dynamics,SI}, 
we calculate the dispersion relations of each system and investigate their behaviors at the low-frequency limit.\\ 
\indent First, we consider metamaterials with the embedded inerter and no other additional mass, as illustrated in Fig.\,\ref{fig:meta}(b). The model has two degrees of freedom in the unit cell: $u^M$ - displacement of mass $M$ on the main chain and $u^b$ - displacement of the point between stiffness $k_b$ and inerter $b$ on the side chain. With the parameters $k_b/K = 1$ and $b/M = 10^6$, the dispersion relation plotted in Fig.\,\ref{fig:meta}(c) shows a band gap as the grey-shaded range near $f=a/\lambda \sim 10^{-4}$. The horizontal axis of normalized wave number $q a/ \ pi$ is shown in logarithmic scale because the band-gap effects happen at very long wavelength. This ultra-low frequency band gap is further demonstrated by transmission attenuation in the steady-state dynamics simulation of a finite chain~\cite{SI}, with results shown in Fig.\,\ref{fig:meta}(d). The band gap's lower edge frequency, $f_\textrm{L}$, is the eigen-frequency of the first band at $q=\pi/a$, as labelled by a red square in Fig.\,\ref{fig:meta}(c). Similarly, the band gap's upper edge frequency, $f_\textrm{U}$, is the eigen-frequency of the second band at $q=0$, as labelled by a red circle in Fig.\,\ref{fig:meta}(c). As good non-dimensionalized measures for comparison purposes, we characterize the band gap by two quantities: (1) The starting dimensionless frequency, $f_\textrm{L}$; and (2) The relative gap size $\Delta f = (f_\textrm{U} - f_\textrm{L}) / f_\textrm{L}$. Fig.\,\ref{fig:meta}(e) shows the numerical results as $f_\textrm{L} = 1.125\times 10^{-4}$ with a relative gap size of $\Delta f \approx 41.4\%$.\\
\indent More generally, we can obtain the analytical equations,
\begin{equation}\label{edges_inerter}
    f_{\textrm{L}} = \frac{1}{2 \pi} \sqrt{\chi_b-\sqrt{\chi_b^2 -4\frac{\kappa_b}{\mu_b}}} \quad \text{and} \quad f_{\textrm{U}} = \frac{1}{2 \pi} \sqrt{\frac{\kappa_b}{\mu_b}},
\end{equation}
where $\chi_b = 2\kappa_b + \kappa_b / (2\mu_b) + 2$, $\kappa_b = k_b/K$, and $\mu_b = b/M$. These closed-form results enable us to perform asymptotic convergence analyses~\cite{SI}. At the limit of large inertance, $\mu_b \gg \kappa_b$, we have
\begin{equation}\label{edges_inerter_large_mu}
    f_{\textrm{L}} \rightarrow \frac{1}{2 \pi} \sqrt{\frac{\kappa_b}{\mu_b (\kappa_b+1)}} \quad \text{and} \quad \Delta f \rightarrow \sqrt{\kappa_b + 1} - 1.
\end{equation}
These equations reveal a unique advantage of the design with embedded inerters: As the inertance $b = \mu_b M$ increases, the band gap shifts to a lower frequency. At the same time, the relative gap size, $\Delta f$, approaches a finite and low limit, keeping the band gap open at very low frequencies. This convergence is also shown together with numerical results in Figs.\,\ref{fig:meta}(e) and \ref{fig:meta}(f) with $\kappa_b=k_b/K=1$, where the gap size converges to $\Delta f = \sqrt{2}-1 \approx 41.4\%$ for large $\mu_b$. In the same limit, we can also get the modal displacement ratios at the gap edges as:
\begin{equation}\label{inerter_large_mu_diaplacement}
\begin{split}
U^b / U^M \rightarrow 1 + 2/\kappa_b \quad &\text{at} \quad f = f_{\textrm{L}},\\
U^b / U^M \rightarrow 1 \quad &\text{at} \quad f = f_{\textrm{U}},
\end{split}
\end{equation}
where $U^b$ and $U^M$ are modal amplitudes of $u^b$ and $u^M$, respectively. Furthermore, taking the additional limit of $\kappa_b \gg 1$ gives
\begin{equation}\label{edges_inerter_final_limit}
\begin{split}
    f_{\textrm{L}} \rightarrow \frac{1}{2 \pi} \sqrt{\frac{1}{\mu_b}}, \quad \Delta f \rightarrow \sqrt{\kappa_b},\\
    U^b / U^M \rightarrow 1 \quad \text{at both } f_{\textrm{L}} \text{ and } f_{\textrm{U}}.
\end{split}
\end{equation}
This shows that it is beneficial to have stiff connections between the inerter and the main chain: While we can get larger gap size as $k_b = \kappa_b K$ increases, the gap's starting frequency will saturate and converges to a finite limit, retaining the ultra-low frequency feature with large inertance. This high-stiffness convergence is shown together with numerical results in Figs.\,\ref{fig:meta}(g) and \ref{fig:meta}(h) with $\mu_b=b/M=10^6$. The lower gap edge converges to $f_{\textrm{L}} = 10^{-3}/(2\pi) \approx 1.59 \times 10^{-4}$.\\ 

\begin{figure}[h!]
\includegraphics[width=0.3\textwidth]{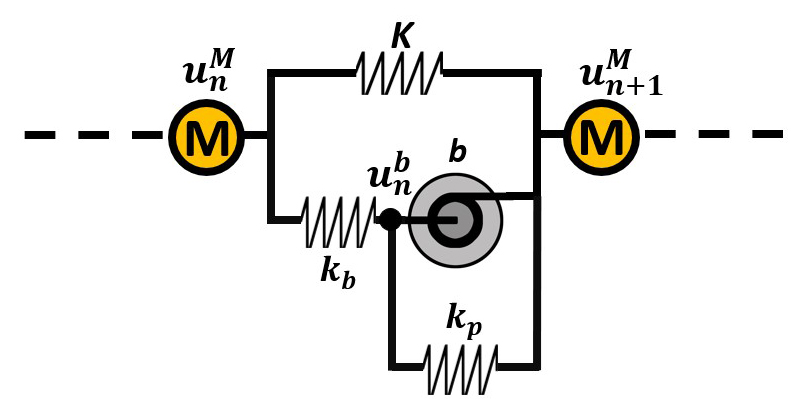}
\caption{\label{fig:kp} The unit cell of Inerter-based metamaterial with an additional stiffness, $k_p$, parallel to the embedded inerter.}
\end{figure}
\indent An important variant design involves an additional stiffness, $k_p$, parallel to the embedded inerter. As shown in Fig.\,\ref{fig:kp}, each unit cell still has two degrees of freedom. We perform the same analyses as before and obtain the equations for both upper and lower edges of the band gap in this case as\\ 
\begin{subequations}
\begin{align}
f_{\textrm{L}} &= \frac{1}{2\pi}\sqrt{\chi_p-\sqrt{\chi_p^2-\frac{4(\kappa_b + \kappa_b \kappa_p + \kappa_p)}{\mu_b}}}\label{kp_type_5_band_lower_5_2},\\
f_{\textrm{U}} &= \frac{1}{2\pi}\sqrt{\frac{(\kappa_b+\kappa_p)}{\mu_b}},\label{kp_type_5_band_upper_5_1}
\end{align}
\end{subequations}
\noindent where $\chi_p = 2\kappa_b + (\kappa_b + \kappa_p)/(2\mu_b) +2$, $\kappa_b =k_b/K$, $\kappa_p =k_p/K$, and $\mu_b = b/M$. At the limit of large inertance $(\mu_b \gg \kappa_b, \kappa_p)$ and taking $\kappa_b = 1$, we get the asymptotic convergence\\
\begin{equation}
    f_\textrm{L} \rightarrow \frac{1}{2\pi}\sqrt{\frac{2\kappa_p + 1}{2\mu_b}} \quad \text{and} \quad
    \Delta f \rightarrow \sqrt{1+\frac{1}{2\kappa_p + 1}}-1,
\end{equation}
which indicate that, as the parallel stiffness $k_p$ increases, not only does $f_{\textrm{L}}$ get higher, but $\Delta f$ also gets smaller. Hence, the parallel stiffness $k_p$ has only detrimental effects, and it is best to set $k_p$ to zero to achieve both design objectives of lower gap frequency and larger gap size.\\ 
\indent Therefore, the optimal design requires that the inerter has very large inertance and very stiff connections with the base structure, and it is better \textit{not} to incorporate any stiffness parallel to the inerter. Because it is possible to fabricate inerters with inertance more than a million times of its actual mass ($\mu_b \sim 10^6$)~\cite{nakaminami2017dynamic}, the embedded inerter design shown in Fig.\,\ref{fig:meta}(b) is practical for engineering applications.\\
Next, we study an alternative metamaterial design with embedded mass-inerter resonators, as illustrated in Fig.\,\ref{fig:para}(a). One end of the inerter is connected to the main chain, while the other end is connected to a resonator mass $m$. This results in a model with three degrees of freedom in each unit cell:  $u^M$ - displacement of mass $M$ on the main chain, $u^b$ - displacement of the point between stiffness $k_b$ and inerter $b$, and $u^m$ - displacement of the resonator mass $m$. Applying the same Bloch-wave procedures as before yields the dispersion bands for this system~\cite{SI}. At the limit of large inertance $(\mu_b \gg \mu_m, \kappa_b, \kappa_f)$ and low frequency $(f \ll 1)$, we arrive at
\begin{equation}\label{edges_inerter_resonator_final_limit}
    f_{\textrm{L}} \rightarrow \frac{1}{2 \pi} \sqrt{\frac{\kappa_b \kappa_m}{\mu_b(\kappa_b + \kappa_m)}} \quad \text{and} \quad \Delta f \rightarrow 0,\\
\end{equation} 
where $\mu_b=b/M$, $\mu_m=m/M$, $\kappa_b=k_b/K$, and $\kappa_f=k_f/K$. Since increasing the inertance closes the band gap ($\Delta f \rightarrow 0$), this mass-inerter-resonator design shown in Fig.\,\ref{fig:para}(a) is not suitable to achieve ultra-low frequency band gaps.\\
 
\begin{figure}[t]
\includegraphics[width=0.5
\textwidth]{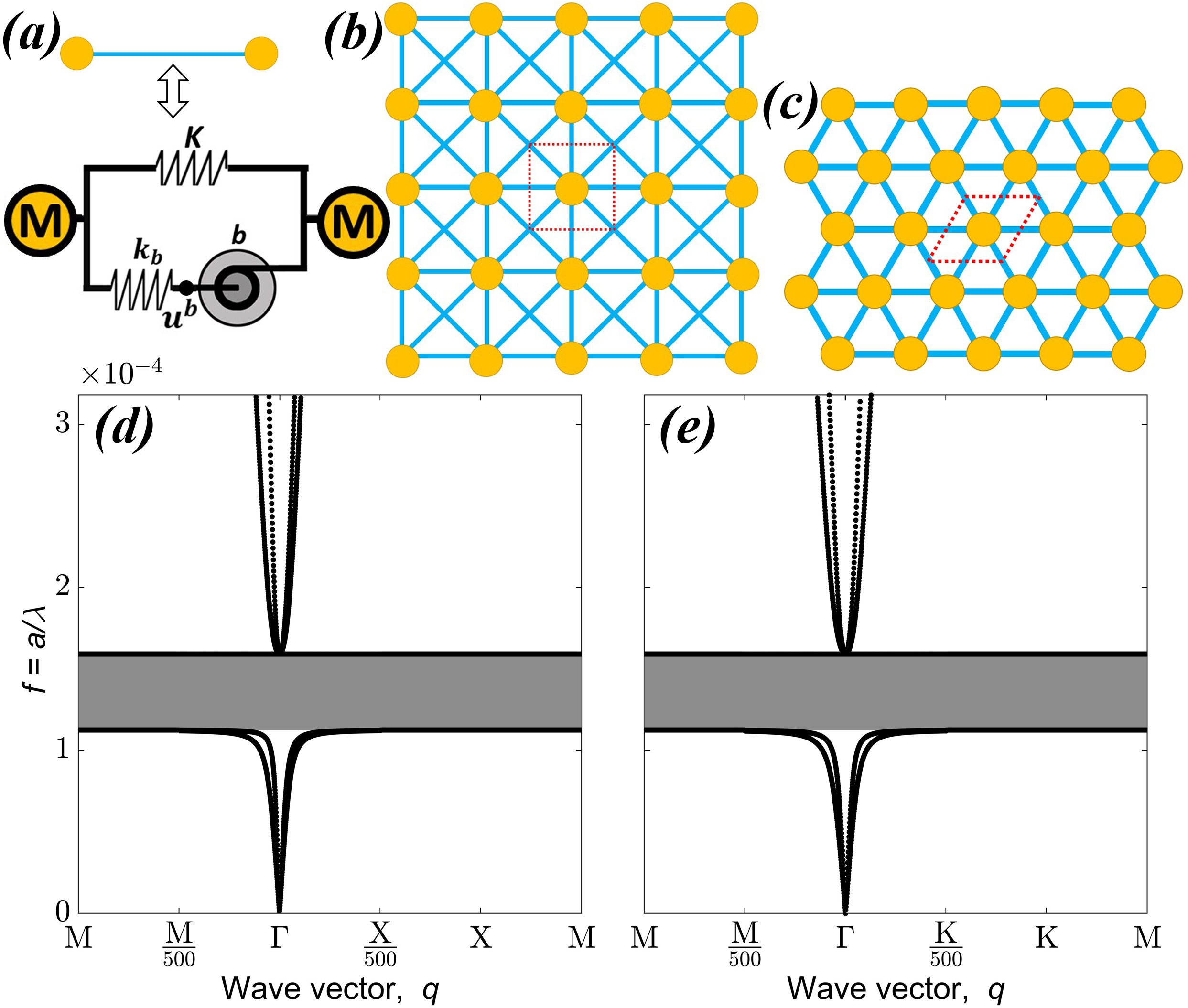}
\caption{\label{figure_stacked_2D} Two-dimensional lattices with embedded inerters: (a) Basic connection unit used in both lattices. Note that $u_b$ here is always just a single degree of freedom in any two-dimensional lattice, as we constrain it to allow displacement parallel to the connection only. 
(b) and (c) depict the square and triangular lattices with embedded inerters, respectively. With $\kappa_b  = k_b/K = 1$ and $\mu_b = b/M  = 10^6$, (d) and (f) show the dispersion bands of configurations (b) and (c), respectively.}
\end{figure}

\indent Lastly, we also look into the traditional locally resonant metamaterials with embedded mass-resonators, as illustrated in Fig.\,\ref{fig:para}(b). The discrete model has two degrees of freedom in each unit cell:  $u^M$ - displacement of mass $M$ on the main chain, and $u^m$ - displacement of the resonator mass $m$. The analytical closed-form expressions of the band gap edges are~\cite{al2017formation,SI}
\begin{subequations}
\begin{align}
f_{\textrm{L}} &= \frac{1}{2 \pi} \sqrt{\chi_m-\sqrt{\chi_m^2 -4\frac{\kappa_m}{\mu_m}}},\label{lower_edges_mass_resonator}\\ 
f_{\textrm{U}} &= \frac{1}{2 \pi} \sqrt{\frac{\kappa_m}{\mu_m} + \kappa_m},\label{upper_edges_mass_resonator}
\end{align}
\end{subequations}
where $\chi_m = \kappa_m/2 + \kappa_m/(2\mu_m) + 2$, $\kappa_m = k_m/K$, and $\mu_m = m/M$. Based on Eq.\,\eqref{lower_edges_mass_resonator}, in order to achieve ultra-low-frequency band gaps with $f_\textrm{L} \sim 10^{-4}$, we need $\mu_m / \kappa_m \sim 10^{8}$. On the other hand, we need to avoid the case of $\mu_m = m/M \gg 1$ since it would make the embedded mass-resonator too heavy as compared to the matrix material or base structure, and hence would be infeasible in most applications. The only viable choice is to adopt the ultra-low-stiffness design with $\mu_m \sim 1$ and $\kappa_m \ll 1$, at which limit we get:
\begin{equation}\label{edges_mass_resonator_limit}
    f_\textrm{L} \rightarrow \frac{1}{2\pi}\sqrt{\frac{ \kappa_m}{\mu_m}} \quad \text{and} \quad
    \Delta f \rightarrow \sqrt{\mu_m + 1} - 1.
\end{equation}
This approach to form a band gap at ultra-low frequencies may initially seem possible. In fact, with $\mu_m=m/M=1$ and $\kappa_m = k_m/K = 5 \times 10^{-7}$, we obtain exactly the same dispersion bands as those plotted in Fig.\,\ref{fig:meta}(c). However, in the same limit of $\mu_m \sim 1$ and $\kappa_m \ll 1$, we get the modal displacement ratios at the gap edges as
\begin{equation}\label{displacement_mass_resonator_limit}
\begin{split}
U^b / U^M \rightarrow 4/\kappa_m \quad &\text{at} \quad f = f_{\textrm{L}},\\
U^b / U^M \rightarrow 1/\mu_m \quad &\text{at} \quad f = f_{\textrm{U}}.
\end{split}
\end{equation}
Hence, the same design parameters give rise to a very high modal displacement ratio, $U^m/U^M \rightarrow 4 / \kappa_m = 8 \times 10^6$
at the lower gap edge $f_\textrm{L}$ (marked with red square in Fig\,\ref{fig:meta}(c)). This means the resonator mass would vibrate with an amplitude that is millions of times of the vibration amplitude in the main chain. Therefore, the ultra-low stiffness design here is impractical in most application scenarios.\\ 
\indent Based on the analyses of all three designs above, we conclude that the inerter-based metamaterial design depicted in Fig.\,\ref{fig:meta}(b) is the only suitable solution to achieve band gaps at the ultra-low dimensionless frequency of $f=a/\lambda \sim 10^{-4}$ or lower. \\
\indent To further demonstrate the efficacy and practicality of this design, we perform numerical studies on two-dimensional lattices with embedded inerters. Due to the close relevance to engineering applications, there have been many pioneering studies realizing band gaps in two- and three-dimensional metamaterials with other lever- or geometry-based inertial amplified structures~\cite{acar2013experimental,taniker2015design,zaccherini2021stress,zaccherini2021granular}, as well as two-dimensional structures that incorporate a flat plate \cite{russillo2022ultra, wang2020tunable, li2019reduction,wang2022sound} or a beam \cite{fang2018band, fang2021inertant, cajic2021tuning,dong2021enhancement,zhou2022low, dong2022energy, aladwani2022tunable} as the base structure with inerter-based resonators. However, none of the designs so far can achieve band gaps at the ultra-low dimensionless frequency of $f=a/\lambda \sim 10^{-4}$. In the following, we investigate the viable design illustrated in Fig.\,\ref{fig:meta}(b) in two-dimensional systems.\\ 
\indent Here, we show conceptual two-dimensional designs in the form of inerter-in-lattice configurations that can  represent various engineering structures. As shown in Fig.\,\ref{figure_stacked_2D}(a), for graphic compactness, we use a blue straight line to represent a connection with main-chain stiffness $K$, side-chain stiffness $k_b$, and side-chain inertance $b$. We assume the displacement $u^b$ at the point between stiffness $k_b$ and inerter $b$ is always rigidly constrained in the lateral direction of the connection unit, so that $u_b$ is always along the direction of the connection. Then, we construct square ( Fig.\,\ref{figure_stacked_2D}(b)) and triangular (Fig.\,\ref{figure_stacked_2D}(c)) lattices with this basic connection building block. All connections possess an embedded inerter on the side chain. Specifically for the square lattice, the diagonal connections are necessary to make the lattice statically stable, so that no zero-frequency band can exist. The crossing point of the two diagonal connections at the center of each square is not a joint, and there is no interactions here between the two diagonal connections. Although the formulations and derivations~\cite{SI} are more challenging than those for one-dimensional cases, we can still study the two-dimensional designs via numerical results. With stiffness ratio $\kappa_b = kb/K = 1$ and inertance ratio $\mu_b = b/M  = 10^6$, we plot the dispersion curves of square and triangular lattices with embedded inerters in Figs.\,\ref{figure_stacked_2D}(d) and \ref{figure_stacked_2D}(e), respectively. A band gap exists in both cases with the lower gap edge frequency of $f_\textrm{L} = 1.125\times 10^{-4}$ and the relative gap size of $\Delta f \approx 41.4\%$. Not only do the band gaps match each other in the two different lattices, but they also match the band gap of the one-dimensional configuration shown in Fig.\,\ref{fig:meta}(b). This shows that the analyses and conclusions on band gaps about one-dimensional models directly extend to higher dimensional scenarios, and the difference in lattice configurations has negligible impact on the ultra-low-frequency band gap. Conceptually, in the very long wavelength limit of $\lambda \gg a$, the unit-cell level geometries and small length-scale details have minimal influence on the metamaterial behavior in the low-frequency regime.\\ 
To conclude, our analytical and numerical analyses offer clear guidelines to design elastic metamaterials with an ultra-low-frequency band gap: Each unit cell needs an embedded inerter with both terminals connected to the base material; no additional resonator mass should be used; to achieve band gaps at lower frequency, higher inertance is needed; and, to achieve wider band gaps, stiffer connection between the inerter and the base material is needed. These insights provide actionable guidelines for future studies towards low-frequency vibration mitigation using metamaterials. Although this study is focused on periodic metamaterials, our analyses can be extended to quasi-crystalline~\cite{Rosa_2021,Thomes_2021,Liu_2022_Topological}, hyper-uniform~\cite{Gkantzounis_2017,Romero_Garc_a_2021,Kuznetsova_2021}, amorphous~\cite{mitchell2018amorphous} or other non-periodic inerter-based metamaterial designs.\\
\indent Both R.G.P and P.W. acknowledge the start-up research funds of the Department of Mechanical Engineering at University of Utah. The support and resources from the Center for High Performance Computing at the University of Utah are gratefully acknowledged. The authors thank Prof.\,Peter Zhu,  Dr.\,Xiaolong Tong, Sharat Paul and Md Nahid Hasan for inspirational discussions.

\normalem
\bibliography{bibliography} 
\end{document}